\documentstyle[preprint,floats,prd,aps]{revtex}
\begin{document}
\preprint{\vbox {\hbox{RIKEN-AF-NP-292}}}

\draft
\title{Possible large direct CP asymmetry in hadronic   $B^\pm\to\pi^\pm\eta'$ decays} \author{Mohammad R. Ahmady$^a$
\footnote{Email: ahmady@riken.go.jp}
, Emi Kou$^b$\footnote{Email: g9870407@edu.cc.ocha.ac.jp} }
\address{
$^a$LINAC Laboratory, The Institute of Physical
and Chemical Research (RIKEN)\\ 2-1 Hirosawa, Wako, Saitama 351-0106,
Japan \\
$^b$ Department of Physics, Ochanomizu University \\
1-1 Otsuka 2, Bunkyo-ku,Tokyo 112-0012, Japan}

\date{July 1998}
\maketitle
\begin{abstract}
We calculate the branching ratio and direct CP asymmetry in nonleptonic two body $B$ decays $B^\pm\to\pi^\pm\eta'$.  It is shown that the tree diagram and gluon fusion mechanism via penguin diagram have comparable contributions to these decays which, as a result, could provide an interesting venue for investigating CP violation. Our estimate shows that the direct CP asymmetry in the above decays could be as large as $75\%$ which along with a branching ratio $B(B^-\to\pi^-\eta')=3.4\times 10^{-6}$ should be accessible to experiment in the near future.
\end{abstract}
%

\newpage

The discovery by the CLEO collaboration of a larger than expected branching ratio for fast $\eta'$ production in hadronic $B$ decays\cite{s,b,bw} has led to extensive theoretical work on investigating the underlying mechanism.  One explanation which is based on the gluon-gluon-$\eta'$ anomalous coupling, has been proposed by Atwood and Soni\cite{as}.  In this mechanism, $\eta'$ is produced by the fragmentation of the virtual gluon of the QCD $b\to sg$ penguin.  On the other hand, in order to formulate the inclusive $B\to X_s\eta'$ and exclusive $B\to K\eta'$ decays under the same mechanism, we proposed a nonspectator gluon fusion process using anomaly driven $g-g-\eta'$ vertex\cite{aks}.  The exclusive branching ratio $BR(B\to K\eta' )$ obtained this way is in good agreement with experimantal data.

In this paper, we focus on a different hadronic decay mode of $B$ mesons which may receive a significant contribution from the above mechanism. 
Indeed, if the anomalous $g-g-\eta'$ vertex in conjunction with the QCD penguin is the underlying process for the fast $\eta'$ production in $B$ meson decays, one could expect that the same mechanism also be an important part of the two body $B^\pm\to\pi^\pm\eta'$ decay modes.  The difference between these decay modes and $B\to K\eta'$ decay is in that, unlike the latter one, the former decays receive a comparable contribution from the tree diagrams.  As a result, one could expect that the above decay modes provide a suitable avenue to investigate the direct CP asymmetry in charged $B$ meson decays.  In this work, we estimate the branching ratio and direct CP asymmetry in $B^\pm\to\pi^\pm\eta'$ decays and point out a possible large CP asymmetry which should be accessible to experiment in the near future.

We start by repeating the derivation of the nonspectator contribution to $B\to K\eta'$ matrix element from reference \cite{aks}.  Using the dominant chromo-electric component of the QCD penguin, we obtain the following effective Hamiltonian corresponding to Figure 1.
\begin{equation}
H_{eff}=iCH(\bar s\gamma_\mu (1-\gamma_5)T^ab)(\bar q\gamma_\sigma T^a q)\frac{1}{p^2}\epsilon^{\mu\sigma\alpha\beta}q_\alpha p_\beta\;\; ,
\end{equation}
where
\begin{equation}
C=\frac{G_F}{\sqrt{2}}\frac{\alpha_s}{2\pi}V_{tb}V_{td}^*[E(x_t)-E(x_c)]\;\; .
\end{equation}
The coefficient function $E$ is defined as
\begin{equation}
\nonumber
E(x_i)=-\frac{2}{3}Lnx_i+\frac{x_i^2(15-16x_i+4x_i^2)}{6{(1-x_i)}^4}Lnx_i+
\frac{x_i(18-11x_i-x_i^2)}{12{(1-x_i)}^3}\;\; ,
\end{equation}
where $x_i=m_i^2/m_W^2$ with $m_i$ being the internal quark mass\cite{il}.  $H$ is the form factor parametrizing the $g-g-\eta'$ vertex  
\begin{equation}
A^{\mu\sigma}(gg\to\eta')=iH(q^2,p^2,m_{\eta'}^2)\delta^{ab}
\epsilon^{\mu\sigma\alpha\beta}q_\alpha p_\beta\;\; .
\end{equation}
Using the decay mode $\psi\to\eta'\gamma$, $H(0,0,m_{\eta'}^2)$ is estimated to be approximately 1.8 GeV$^{-1}$\cite{as}.  A re-arrangement of Eq. (1) via Fierz transformation
\begin{eqnarray}
\nonumber
(\bar s\gamma_\mu (1-\gamma_5)T^ab)(\bar q\gamma_\sigma T^aq)=\frac{1}{9}
&&\left [(\bar s\gamma_\sigma\gamma_\rho\gamma_\mu (1-\gamma_5)q)(\bar q\gamma^\rho (1-\gamma_5)b)
+(\bar s\gamma_\sigma\gamma_\mu (1+\gamma_5)q)(\bar q(1-\gamma_5)b)\right . \\
&&\left . -\frac{1}{2}(\bar s\gamma_\sigma\sigma_{\rho\eta}\gamma_\mu q)(\bar q\sigma^{\rho\eta}(1-\gamma_5)b) +{\rm color\;\; octect}\right ]\;\; ,
\end{eqnarray}
and using the definition of the decay constants for the $B$ and $K$ mesons in the context of factorization, we can express the matrix element for $B\to K\eta'$ decay as the following:
\begin{equation}
<\eta' K|H_{eff}|B>=-i\frac{2CHf_Bf_K}{9p^2}\left (p_B.qp_K.p-p_B.pp_K.q\right )\;\; ,
\end{equation}
leading to the exclusive decay rate
\begin{equation}
\Gamma (B\to K\eta' )=\frac{C^2H^2f_B^2f_K^2}{486\pi p^4}{\vert \vec p_K\vert}^3\left (3p_0^2{\vert \vec p_K\vert}^2 +(m_{\eta'}^2+{\vert \vec p_K\vert}^2)(p_0^2-p^2) \right )\;\; ,
\end{equation}
where $\vert\vec p_K\vert$ and $p_0$ are the three momentum of the K meson 
\begin{equation}
\vert \vec p_K\vert ={\left [\frac{{(m_B^2+m_K^2-m_{\eta'}^2)}^2}{4m_B^2}-m_K^2\right ]}^{\frac{1}{2}}\;\; ,
\end{equation}
and the energy transfer by the gluon emitted from the light quark in the B meson rest frame,  respectively.  Inserting $p^2\approx -\Lambda_{\rm QCD}^2\approx -0.3^2$ GeV$^2$, $p_0=0.3$ GeV and $f_B=0.2$ GeV in Eq. (7) results in the following exclusive branching ratio:
\begin{equation}
B(B\to K\eta' )=7.1\times 10^{-5}\;\; ,
\end{equation}
which is in good agreement with the experimental data\cite{s,b}.  On the other hand, we observe that if $g-g-\eta'$ anomalous coupling along with QCD $b\to sg$ penguin (Fig. 1) is indeed the dominant underlying mechanism for $B\to K \eta'$ decay, then one could expect that a similar process, with $d$ quark replacing the strange quark, to contribute significantly to $B\to \pi\eta'$ decay mode.  However, in this case, unlike $B\to K\eta'$ decay, the tree and penguin amplitudes receive a CKM suppression factor of the same order and therefore there is a good possibility for existence of a large direct CP asymmetry in $B^\pm\to\pi^\pm\eta'$ decay modes. 
The nonspectator contribution to $B\to\pi\eta'$ can be obtained from Eq. (6) by replacing $f_K$ with $f_\pi$ and $C$ with $C'$, where
\begin{equation}
C'=\frac{G_F}{\sqrt{2}}\frac{\alpha_s}{\pi}\left ( V_{tb}V^*_{td}[E(x_t)-E(x_c)]+V_{ub}V_{ud}^*[E(x_u)-E(x_c)]\right )
\;\; .
\end{equation}
Thus, including the expression for the tree amplitude (Fig. 2) which are derived by using the factorization assumption, the total matrix element for $B^-\to\pi^-\eta'$ can be written as:
\begin{eqnarray}
<\eta' \pi^-|H_{eff}|B^-> &=& -i\frac{C'Hf_Bf_\pi}{9p^2}\left (p_B.qp_\pi.p-p_B.pp_\pi.q\right ) \\
\nonumber
&+&i\frac{G_F}{\sqrt{2}}V_{ub}V_{ud}^*\left (a_1f_\pi F_0^{B\to\eta'}(m_\pi^2)(m_B^2-m_{\eta'}^2)+a_2f_{\eta'}^u F_0^{B\to\pi}(m_{\eta'}^2)(m_B^2-m_\pi^2)\right )\;\; .
\end{eqnarray} 
$a_1=1.03$ and $a_2=0.12$ are the combinations of the Wilson coefficients of the current-current operators and $f_{\eta'}^u$ and $F_0^{B\to P},\; P=\eta',\pi$, are defined as follows:
\begin{equation}
<\eta'(p_{\eta'})|\bar u\gamma^\mu (1-\gamma_5) u|0>=if_{\eta'}^up^\mu_{\eta'}\;\; ,
\end{equation} 
\begin{eqnarray}
\nonumber
<P(p_{P})|\bar u\gamma^\mu (1-\gamma_5) u|B(p_B)> &=& \left [{(p_B+p_{P})}^\mu-\frac{m_B^2-m_P^2}{q^2}q^\mu\right ]F_1^{B\to P}(q^2) \\
 &+& \frac{m_B^2-m_P^2}{q^2}q^\mu F_0^{B\to P}(q^2)
\;\; ,
\end{eqnarray}
where $q=p_B-p_P$.  We estimate $f^u_{\eta'}\approx 0.064$ GeV by using the two-angle formalism for the $\eta -\eta'$ mixing\cite{l,slpt,fk}.  As for the form factors, we use the numerical results $F_0^{B\to\pi}(m_{\eta'}^2)\approx F_0^{B\to\pi}(0)\approx 0.33$ and $F_0^{B\to\eta'}(m_{\pi}^2)\approx F_0^{B\to\eta'}(0)\approx 0.135$ obtained in the BSW model\cite{bsw}.  Using the Wolfenstein parametrization\cite{w} and the unitarity triangle convention for the phases of the quark mixing matrix elements\cite{js,akl}
\begin{equation}
 V_{td}=A\lambda^3(1-\rho -i\eta )=\vert V_{td}\vert e^{-i\beta}\;\;,\;\; V_{ub}=A\lambda^3(\rho-i\eta )=\vert V_{ub}\vert e^{-i\gamma}\;\; ,
\end{equation}
where $A=0.81\pm 0.06$ and $\lambda ={\rm Sin}\theta_c=0.2205\pm 0.0018$ are well determined\cite{pdg}.  The parameters $\rho$ and $\eta$ are correlated and their range have been determined from CKM unitarity fits\cite{al}.  In order to show the dependence of the branching ratio and CP asymmetry on these parameters, we present our results for central values of $A$ and $\lambda$ and three different sets of $(\rho\; , \eta )$ values from the constrained range.

We note that different weak (CP odd) phases enter nonspectator (penguin) and tree amplitudes.  Also, an overall strong (CP even) phase $\theta_s$  for the nonspectator amplitude is assumed which in combination with the weak phases leads to CP asymmerty in $B^\pm\to\pi^\pm\eta'$ decays.  In fact, one possible source of this strong phase could be the form factor which parametrizes the  $g-g-\eta'$ vertex.  For example, in the sigma model formulation of $H(q^2,0,m_{\eta'}^2)$ where this form factor is derived from a quark triangle loop (Fig. 3), an absorptive part is generated for $q^2\ge 4m_f^2$ ($m_f$ is the mass of the quarks participating in the triangle loop i.e., $f=u\; ,\; d\; ,\; s$ in the  $\eta'$ case)\cite{kp,aek}.  For $q^2=2$GeV$^2$ the magnitude of this strong phase can be around $-50^\circ$ to $-60^\circ$ depending on the input value for $m_f$.

Inserting $H=\vert H\vert e^{i\theta_s}$ ($\vert H\vert\approx 1.8$GeV$^{-1}$ as mentioned before) in Eq. (11) and using the same input values for various parameters as in $B\to K\eta'$ case, we obtain the following branching ratio for $B^-\to\pi^-\eta'$
\begin{eqnarray}
B(B^-\to\pi^-\eta' )=   
\nonumber
1.26\times 10^{-6}{\left (\frac{\vert V_{ub}\vert}{0.003}\right )}^2  [1+&0.49&(1+0.42\frac{{\vert V_{td}\vert}^2}{{\vert V_{ub}\vert}^2}-1.3\frac{\vert V_{td}\vert}{\vert V_{ub}\vert}{\rm Cos}(\beta +\gamma )) \\
  &+& 1.06({\rm Cos}(\theta_s)-0.65\frac{\vert V_{td}\vert}{\vert V_{ub}\vert}{\rm Cos}(\theta_s+\beta +\gamma ))  ]\;\; .
\end{eqnarray}
The first term in Eq. (15) is due to current-current operators (tree) while the second term is the contribution of the nonspectator gluon fusion process (penguin) and the last entry is the cross term.  The branching ratio for the CP conjugate process $B^+\to\pi^+\eta'$ is obtained from the Eq. (15) by changing the sign of the weak angles $\beta$ and $\gamma$.  Consequently, the CP asymmetry  which is defined as
\begin{equation}
A_{cp}=\frac{\Gamma (B^-\to\pi^-\eta')-\Gamma (B^+\to\pi^+\eta')}{\Gamma (B^-\to\pi^-\eta')+\Gamma (B^+\to\pi^+\eta')}\;\; ,
\end{equation}
is proportional to ${\rm Sin}(\beta +\gamma)={\rm Sin}(\alpha)$ and ${\rm Sin}(\theta_s)$.  In figure 4, we illustrate the variation of $B(B^-\to\pi^-\eta')$ and $A_{CP}$ when $\theta_s$ takes on values in the range $(-\pi\; ,\; \pi)$ for three different sets of $\rho$ and $\eta$.  It is quite interesting that a large asymmetry is possible due to the comparable nonspectator and tree contributions to this process.  For example, for $\rho =0.25$ and $\eta =0.20$ (i.e. $\vert V_{ub}\vert =0.0028$) , $A_{cp}$ could be as large as $75\%$ and $B(B^-\to\pi^-\eta')=3.4\times 10^{-6}$ if $\theta_s=90^\circ$.  A larger branching ratio $B(B^-\to\pi^-\eta')=1.2\times 10^{-5}$ is possible for $\rho\approx 0$ and $\eta=0.46$ (i.e. $\vert V_{ub}\vert =0.0040$) however, the asymmetry is somewhat smaller around $20\%$.  On the other hand, for the preferred values $\rho =0.12$ and $\eta =0.34$ (i.e. $\vert V_{ub}\vert =0.0031$), the resulting branching ratio and asymmetry are
\begin{equation}
B(B^-\to\pi^-\eta')=2.9\times 10^{-6}\;\; ,\;\; A_{cp}=-0.41\;\; ,
\end{equation}
for $\theta_s\approx -55^\circ$, which is the case if we assume that the absorptive part of the gluon-gluon-$\eta'$ form factor $H$ is responsible for the strong phase.  

In conclusion, we emphasize that a comparison between our results and other theoretical estimates\cite{akl1,akl2} indicates that $B(B^\pm\to\pi^\pm\eta')$ and $A_{CP}$ could receive a significant enhancement due to the inclusion of the nonspectator gluon fusion mechanism. 
These decay modes could be accessible to experiment in the near future and therefore, provide a clean testing ground for CP violation mechanism in the Standard Model.

\section{acknowledgement}
The authors thank Professors I. A. Sanda and A. Soni for useful discussions.  M. A. acknowledges support from the Science and Technology Agency of Japan.  E. K. acknowledges support from the Japanese Society for the Promotion of Science.

\newpage
{\center \bf \huge Figure Captions}
\vskip 3.0cm
\noindent
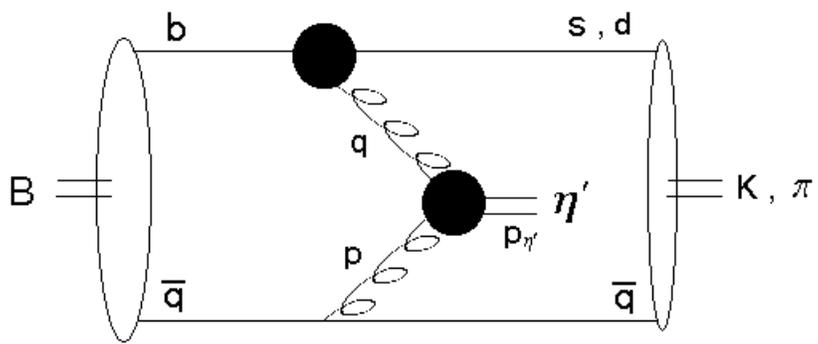
{\bf Figure 1}: Non-spectator contribution to $B\to (K\; ,\;\pi)\eta'$ decays. \\
\vskip 0.5cm
\noindent
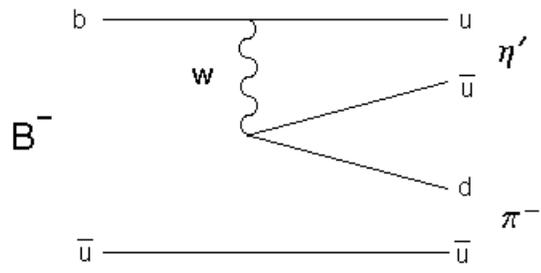
{\bf Figure 2}: The current-current (tree) contribution to $B^-\to\pi^-\eta'$ decay. \\
\vskip 0.5cm
\noindent
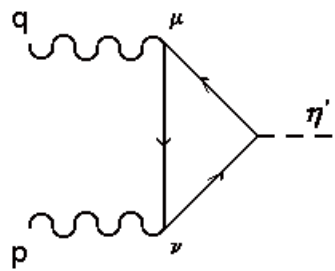
{\bf Figure 3}: The triangle quark loop diagram for gluon-gluon-$\eta'$ vertex. \\
\vskip 0.5cm
\noindent
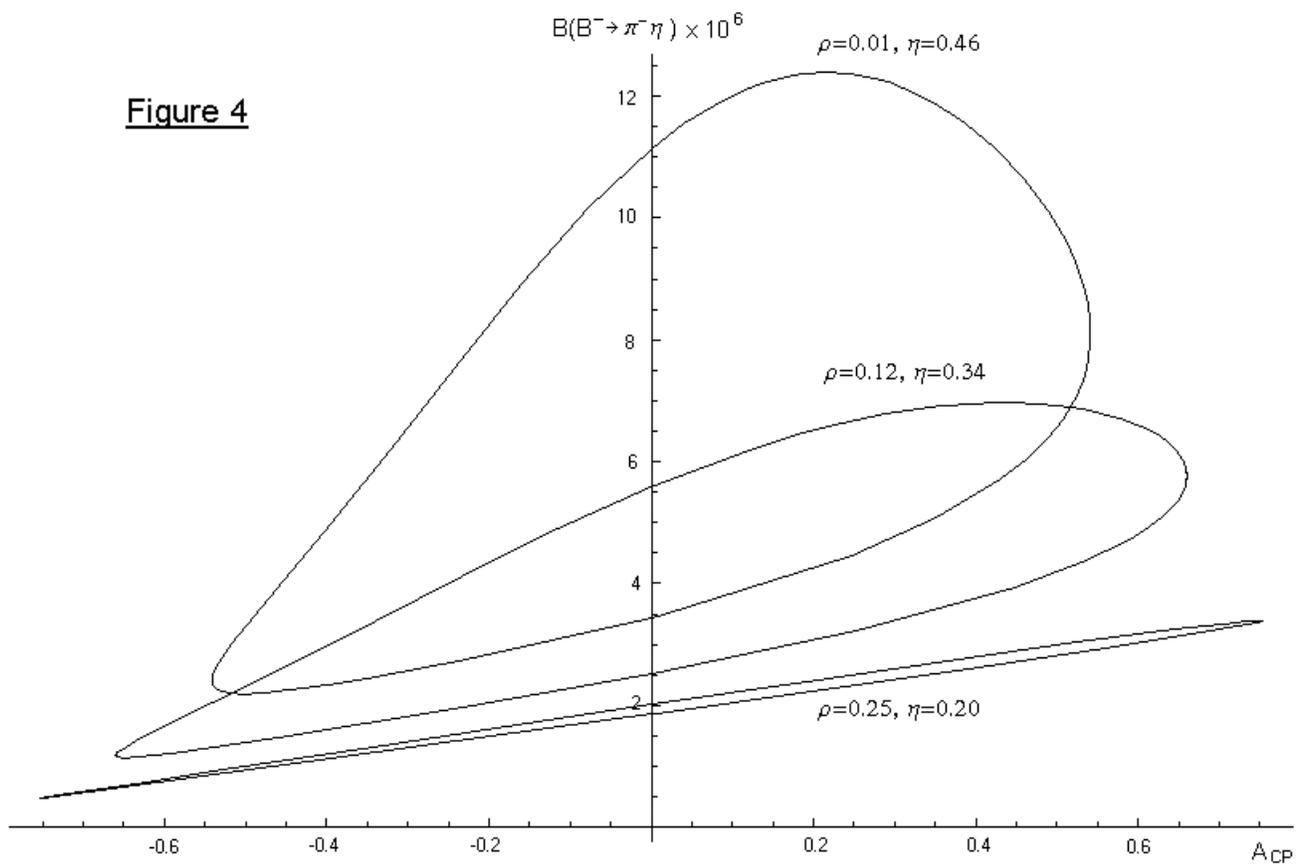
{\bf Figure 4}: Branching ratio versus CP asymmetry as $\theta_s$ varies from $-\pi$ to $\pi$ for three sets of $(\rho\; ,\; \eta )$ values. \\
\vskip 0.5cm

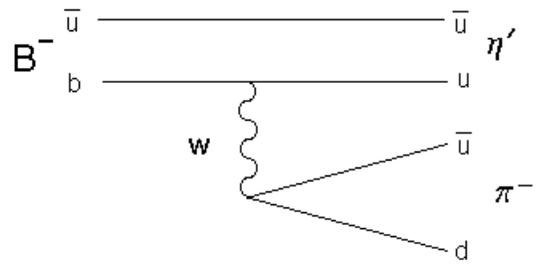
\end{document}